\documentclass[aps,prb,twocolumn,floatfix,showpacs,superscriptaddress]{revtex4}

\usepackage{graphicx}
\usepackage{bbm}
\usepackage{amsmath,amssymb}
\newcommand{\be}{\begin{equation}}
\newcommand{\beq}{\begin{equation}}
\newcommand{\ee}{\end{equation}}
\newcommand{\bea}{\begin{eqnarray}}
\newcommand{\eea}{\end{eqnarray}}
\newcommand{\ba}{\begin{array}}
\newcommand{\ea}{\end{array}}

\renewcommand{\vr} {{\bf r}}

\newcommand{\vf} {{\bf f}}

\begin{document}
\title{Strictly correlated uniform electron droplets}
\author{E. R{\"a}s{\"a}nen}
\email[Electronic address:\;]{esa.rasanen@jyu.fi}
\affiliation{Nanoscience Center, Department of Physics, 
University of Jyv{\"a}skyl{\"a}, FI-40014 Jyv{\"a}skyl{\"a}, Finland}
\author{M. Seidl}
\affiliation{Institute of Theoretical Physics, University of 
Regensburg, D-93040 Regensburg, Germany}
\author{P. Gori-Giorgi}
\affiliation{Department of Theoretical Chemistry and Amsterdam Center for Multiscale Modeling, FEW, Vrije Universiteit, De Boelelaan 1083, 1081HV Amsterdam, The Netherlands}


\begin{abstract}
We study the energetic properties of finite but internally
homogeneous $D$-dimensional electron droplets in the 
strict-correlation limit.
The indirect Coulomb interaction is found to increase as
a function of the electron number, approaching the
tighter forms of the Lieb-Oxford bound recently proposed by R{\"a}s{\"a}nen {\it et al.} 
[Phys. Rev. Lett. {\bf 102}, 206406 (2009)].
The bound is satisfied in three-, two-, and one-dimensional 
droplets, and in the latter case it is reached
{\em exactly} -- regardless of the type of interaction considered.
Our results provide useful reference data for 
delocalized strongly correlated systems, and
they can be used in the development and testing of 
exchange-correlation density functionals
in the framework of density-functional theory.
\end{abstract}


\maketitle

\section{Introduction}

Strongly correlated materials have 
attracted tremendous interest across 
different fields of physics.~\cite{QuiHoo-PW-09} 
Famous examples of strongly correlated
systems are high-temperature superconductors,
organic conductors, ultracold atoms, 
and semiconductor quantum dots. These
systems provide a particular challenge
to theorists -- simply because their
properties cannot be predicted from the
behavior of individual particles. 

A physically important quantity in
a quantum system is the magnitude
of the indirect particle-particle
interaction. This corresponds to the
energy difference between the expectation
value of the quantum mechanical 
interaction operator and the 
classical interaction energy 
of charged particles [see Eq.~(\ref{w}) below].
Lieb~\cite{Lie-PLA-79} showed that this quantity
has a rigorous lower bound for Coulomb-interacting three-dimensional (3D) 
systems. 
Later on, the bound was tightened~\cite{LieOxf-IJQC-81,ChaHan-PRA-99} and extended to
two-dimensional~\cite{LieSolYng-PRB-95} (2D) systems. In density functional theory (DFT), 
this bound has been extensively used for building and testing approximations for the 
exchange-correlation energy functional 
(see, e.g., Refs.~\onlinecite{Per-INC-91,LevPer-PRB-93,OdaCap-PRA-09} and references therein). 

More recently, using physical rather than formal arguments, an even 
tighter bound for 3D and 2D systems has been proposed together 
with an extension to 
one-dimensional (1D) systems.\cite{RasPitCapPro-PRL-09} 
The basic idea of Ref.~\onlinecite{RasPitCapPro-PRL-09} is that
the tightest form of the lower bound on the indirect interaction 
in $D$ dimensions should correspond to the amount of correlation
in the infinite $D$-dimensional homogeneous electron gas (HEG) 
in the low-density limit.~\cite{RasPitCapPro-PRL-09,Per-INC-91} 
This physically appealing idea provides an improved bound for 3D systems, the
introduction of a relatively tighter bound in 2D, and a 
proposal for the bound in 1D.

Odashima and Capelle\cite{OdaCap-JCP-07} have shown through extensive 
numerical studies that {\em finite} electronic systems
are energetically far above the lower bound, even when considering the 
tighter form of Ref.~\onlinecite{RasPitCapPro-PRL-09}. This has triggered our
interest to construct a finite, yet physically simple 
system that is as close as possible to the bound of 
Ref.~\onlinecite{RasPitCapPro-PRL-09}, or,
if possible, even below (which would imply violation 
of the proposed lower bound).
To challenge the bound maximally for a given density,
the strict-interaction limit of DFT provides a 
suitable methodology. The mathematical structure of this approach
-- corresponding to a system with a given density and maximum
spatial correlation between the electrons -- has been
uncovered in the last three years.\cite{SeiGorSav-PRA-07,GorVigSei-JCTC-09,GorSeiVig-PRL-09} 
Consequently, explicit solutions, at least for centrally symmetric densities,
 have started to become available.\cite{GorSeiSav-PCCP-08,GorSei-PCCP-10} 

In this paper we use the strong-interaction 
limit of DFT to investigate the Lieb-Oxford bound in 3D, 2D, and 1D.
We take the simplest imaginable test system, i.e., a finite $D$-dimensional 
electron droplet of a {\em uniform} density (up to a certain radius 
above which the density is rigorously zero)
and examine its properties as the number of particles changes. 
Our analytic and numerical results show 
no evidence of violation of the
lower bounds proposed in 
Ref.~\onlinecite{RasPitCapPro-PRL-09}, but in all dimensions the 
low-density result of the HEG is
approached as a function of the electron number $N$. 
In 1D, our large-$N$ limit {\em exactly} 
corresponds to the proposed lower bound
-- regardless of the type of electron-electron interaction
examined (contact, soft-Coulomb, and regularized).
In 2D and 3D, on the other hand, the large-$N$ result is
$\sim 2\%$ off the bound, although this small difference is 
within the errors associated with our numerical procedure.

\section{Theory}
\label{sec_theory}

\subsection{Lower bound on the indirect interaction}\label{lower}

We consider a system of interacting electrons described
by the Hamiltonian 
\be
{\hat H}={\hat T}+{\hat V}_{ee}+{\hat V}_{\rm ext},
\ee
where ${\hat T}$ is the kinetic-energy operator,
${\hat V}_{ee}$ 
is the electron-electron (e-e) interaction, 
and ${\hat V}_{\rm ext}$ accounts for an
external local one-body potential. We can define the {\em indirect} 
(quantum mechanical) part of the e-e interaction as
\be\label{w}
\widetilde{W}[\Psi]\equiv \left<\Psi|{\hat V}_{ee}|\Psi\right>-U[n_\Psi],
\ee
where 
\be
U[n]=\frac{1}{2}\int d{\vr'} \int d{\vr}\,n(\vr) n(\vr') V_{ee}(|\vr-\vr'|)
\ee
is the classical (Hartree) interaction calculated from the (charge)
density $n(\vr)$. In Eq.~(\ref{w}), 
$\Psi=\Psi(\vr_1\sigma_1,\ldots,\vr_N\sigma_N)$
is an arbitrary $N$-electron wavefunction (where $\sigma_i$ denote
spin variables) and $n_\Psi(\vr)$ is the density associated with it.
The indirect e-e interaction has an important lower bound which can be
expressed as
\be
\widetilde{W}[\Psi]\geq - \; C_D \int d^D r\,n_\Psi^{1/D+1}(\vr),
\label{bound}
\ee
where $D=3,2,1$ is the dimension.
In 3D, the bound originally found by Lieb~\cite{Lie-PLA-79} 
is best known as the 
Lieb-Oxford (LO) bound~\cite{LieOxf-IJQC-81} having a prefactor 
$C_3^{\rm LO}=1.68$. The bound has been tightened
by physical, yet nonrigorous arguments to $C_3=1.44$ 
(Ref.~\onlinecite{RasPitCapPro-PRL-09}).
In 2D the existence of the bound was proven by Lieb, Solovej, and 
Yngvason~\cite{LieSolYng-PRB-95}
(LSY) with $C_2^{\rm LSY}=192\sqrt{2\pi}\approx 481.28$.
In Ref.~\onlinecite{RasPitCapPro-PRL-09}, a
tighter bound of $C_2=1.96$ was proposed.

The bound in Eq.~(\ref{bound}) was originally constructed for
Coulomb-interacting systems, where ${\hat V}_{ee}=\sum_{i<j}|\vr_i-\vr_j|^{-1}$.
In 1D, however, this type of interaction is ill-defined due to 
the divergence at $|x_i-x_j|=0$. In Ref.~\onlinecite{RasPitCapPro-PRL-09} it
was shown that a 1D bound can be constructed by applying
a contact interaction or a soft-Coulomb interaction. In Sec.~\ref{1D} below,
the 1D case is studied in detail considering three types of
the e-e interaction.

The bound of Eq.~(\ref{bound}) can be equivalently 
expressed as\cite{Per-INC-91,LevPer-PRB-93,OdaCap-JCP-07,OdaCap-IJQC-08} 
\be
\lambda_D[\Psi]\equiv\frac{\widetilde{W}[\Psi]}{E_x^{\rm LDA}[n_\Psi]}\le \frac{C_D}{A_D}\equiv{\bar \lambda}_D.
\label{eq_boundonlambda}
\ee
where
\be
E_x^{\rm LDA}[n]=-A_D\int d^D r\,n^{1/D+1}(\vr)
\label{LDA}
\ee
is the local density approximation (LDA) 
for the electronic exchange energy, corresponding to the exact 
exchange energy for the HEG. Here the prefactors are
given by $A_3=3^{4/3}\pi^{-1/3}/4$ and $A_2=2^{5/2}\pi^{-1/2}/3$ 
(for the 1D case see Sec.~\ref{1D}).
In the right-hand side of Eq.~(\ref{eq_boundonlambda}),
the values obtained for ${\bar \lambda}_D$
in 3D, 2D, and 1D are 
\be
\bar{\lambda}_3=1.96,\qquad    \bar{\lambda}_2=1.84,\qquad
     \bar{\lambda}_1=2.
\ee
They have been proposed as the {\em tightest} bounds with the 
prefactors $C_D$ given above -- hence the bar symbol to differentiate 
from the functional $\lambda_D[\Psi]$.
The upper bounds ${\bar \lambda}_D$ correspond to the 
low-density limit of the $D$-dimensional HEG. 
The physical argumentation~\cite{RasPitCapPro-PRL-09}
behind the HEG result was motivated by the finding of
Lieb and Oxford, who showed that there is a 
function $\tilde{\lambda}_3(N)$ which provides an upper bound for all systems 
with particle number equal to $N$ (Ref.~\onlinecite{LieOxf-IJQC-81,OdaCapTri-JCTC-09}).
The function $\tilde{\lambda}_3(N)$ is monotonic, 
with $\tilde{\lambda}_3(N+1)\ge \tilde{\lambda}_3(N)$, so that the most 
general bound of Eq.~(\ref{eq_boundonlambda}) is obtained by 
considering $N\to\infty$.

In this paper we focus on the question how
the LO bound can be {\em challenged}.
In other words, how the wavefunction $\Psi$ in 
Eq.~(\ref{eq_boundonlambda}) must be chosen such 
that $\lambda_D[\Psi]$ becomes as large as possible? 
For any class of wavefunctions with a given fixed 
density $n(\vr)$ (Ref.~\onlinecite{Lev-PNAS-79}) the 
answer to this question is
\be
\max_{\Psi\to n}\lambda_D[\Psi]\equiv \Lambda_D[n]\equiv\frac{W_\infty[n]}{E_x^{\rm LDA}[n]},
\label{eq_Lambda}
\ee
where
\be
W_\infty[n]\equiv\min_{\Psi\to n}\widetilde{W}[\Psi]=\min_{\Psi\to n}\langle\Psi|\hat{V}_{ee}|\Psi\rangle-U[n],
\label{eq_Winfty}
\ee
is the indirect Coulomb energy in the strong-interaction limit of DFT, which can be now calculated (at least for centrally symmetric densities) with the theory of strictly correlated electrons.\cite{SeiGorSav-PRA-07}
This quantity was also considered in the original proof of the bound.\cite{Lie-PLA-79,LieOxf-IJQC-81,Per-INC-91,LevPer-PRB-93} 
In the following section we briefly review 
how the functional $W_\infty[n]$ of Eq.~(\ref{eq_Winfty}) 
can be constructed for a given density $n(\vr)$.

\subsection{Strong-interaction limit}
\label{sec_SCE}
We may define $\Psi^\alpha[n]$ as the wavefunction that
minimizes $\big<\Psi^\alpha|{\hat T}+\alpha{\hat V}_{ee}|\Psi^\alpha\big>$ --
corresponding to a system where the interaction is scaled -- 
with the constraint of reproducing the given density $n(\vr)$. 
The scaled indirect Coulomb interaction  
$W_\alpha[n]=\langle \Psi^\alpha[n]|\hat{V}_{ee}|\Psi^\alpha[n]\rangle - U[n]\equiv\langle V_{ee}^\alpha\rangle - U[n]$ satisfies a set of useful
exact relations within DFT.~\cite{LevPer-PRA-85} 

Here we consider
the strong-interaction limit $\alpha\rightarrow\infty$,
where it is sufficient to minimize the interaction term
alone, since $\big<\alpha{\hat V}_{ee}\big>$ grows
faster than $\big<{\hat T}\big>\propto\alpha^{1/2}$ (Refs.~\onlinecite{Sei-PRA-99,GorVigSei-JCTC-09}). As anticipated in Eq.~(\ref{eq_Winfty}), we thus compute explicitly, for a given density $n(\vr)$, the functional
\be
\big<{\hat V}^\infty_{ee}\big>\equiv \min_{\Psi\to n}\langle \Psi|\hat{V}_{ee}|\Psi\rangle.
\label{eq_theSCEfunctional}
\ee
In the strong-interaction limit of DFT the electrons minimize their interaction energy 
while reproducing the given {\em smooth} density $n(\vr)$. This $\alpha\to\infty$ limit 
is thus different from the more commonly considered Wigner crystal,~\cite{GiuVig-BOOK-05} 
as here the one-electron density is fixed {\em a priori} (and can be very different from 
the one of a Wigner-like structure, e.g., it can be the density of a weakly correlated 
system like a neutral atom\cite{SeiGorSav-PRA-07}). As discussed in detail in 
Refs.~\onlinecite{SeiGorSav-PRA-07,GorVigSei-JCTC-09,GorSei-PCCP-10}, in the $\alpha\to\infty$ 
limit the relative positions of the electrons become strictly correlated: the position
$\vr_1=\vr$ of the first electron determines the positions $\vr_i$ of all
the other electrons via $N-1$ {\em co-motion functions} $\vf_i(\vr)$, $\vr_i=\vf_i(\vr)$. Thus, the
probability of finding the first electron in the volume element 
$d\vr$ around the position $\vr$ is the same as finding the 
$i^{\rm th}$ electron in the volume element $d\vf_i(\vr)$ around $\vf_i(\vr)$, so that the co-motion functions are linked to the density through the differential equation $n(\vr) d \vr=n(\vf_i(\vr)) d \vf_i(\vr)$, which has to be solved with the boundary condition that the corresponding
expectation value of the interaction operator,\cite{SeiGorSav-PRA-07}
\be
\big<{\hat V}^\infty_{ee}\big>=W_\infty[n]+U[n]=\sum_{i=1}^{N-1}\sum_{j=i+1}^{N}\int d\vr\,\frac{n(\vr)/N}{|\vf_i(\vr)-\vf_j(\vr)|},
\label{eq_VeeSCE}
\ee
is minimum.

\begin{figure}
\includegraphics[width=0.8\columnwidth]{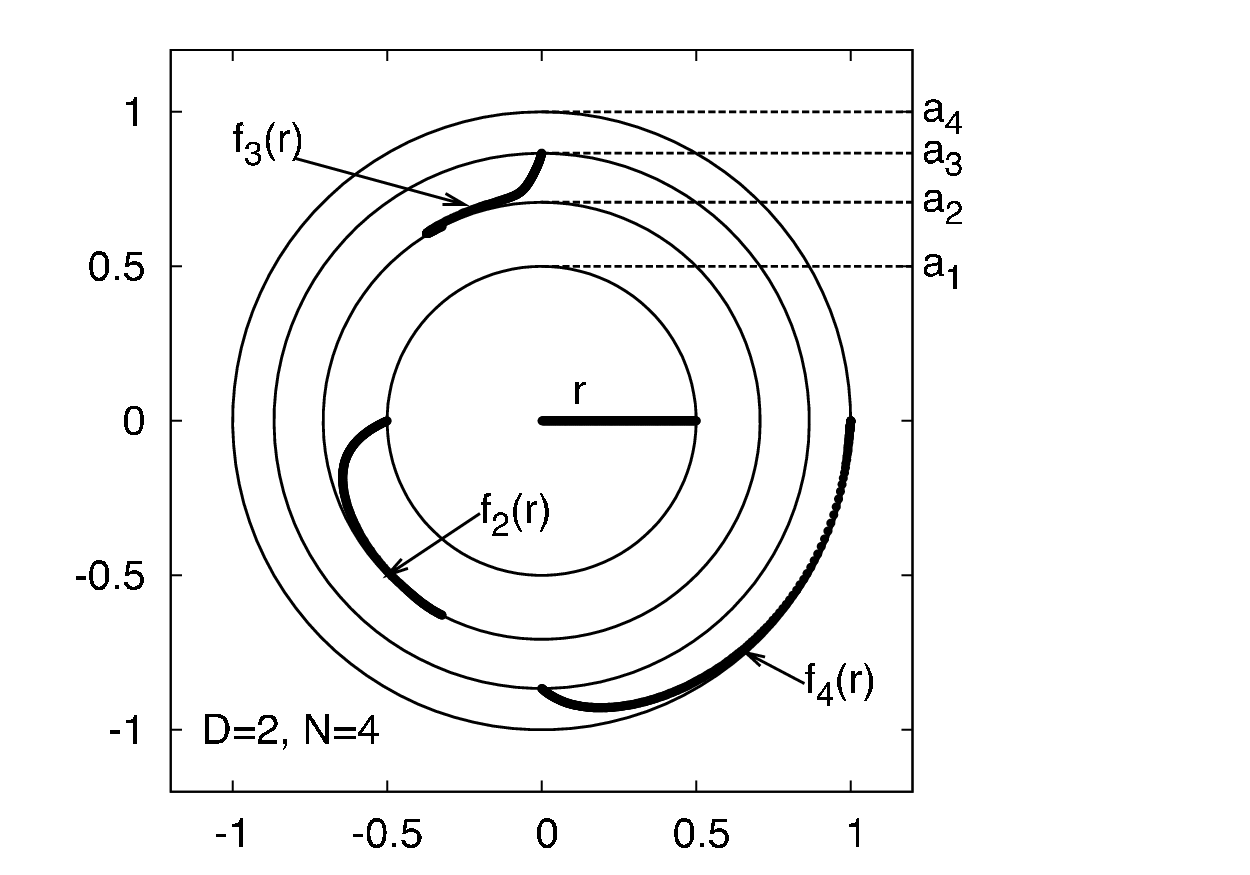}
\caption{Example of a section of the vectorial co-motion functions $\vf_i(r)$ for a uniform droplet in two dimensions with $N=4$ electrons. As the position of the first electron changes on the $x$-axis from $x=0$ to $x=a_1=N_e^{-1}(1)$, the position of the second electron changes along the boldface curve from $r=a_2=N_e^{-1}(2)$ to $r=a_1$, the position of the third electron from $r=a_2$ to $r=a_3=N_e^{-1}(3)$, and the one of the fourth electron from $r=a_4=N_e^{-1}(4)$ to $r=a_3$.}
\label{fig_positionsD2N4}
\end{figure}

Following Refs.~\onlinecite{SeiGorSav-PRA-07,GorSei-PCCP-10}, we consider here
a spherical (circular) density in 3D and 2D, for which the $\vf_i(\vr)$ can be constructed as follows. Given the expected number of electrons between $0$ and $r$,
\be
N_e(r)=\int_0^r dr'\,S(r')\,n(r'),
\ee
where $S(r)=4\pi r^2$ and $2\pi r$ in
3D and 2D, respectively, 
the general solution for the radial co-motion functions
in a centrally symmetric $N$-electron system can be written as\cite{SeiGorSav-PRA-07}
\be
f_{2k}(r)=
\Bigg\{
\begin{array}{c}
N_e^{-1}[2k-N_e(r)],\; r\leq a_{2k} \\
N_e^{-1}[N_e(r)-2k],\; r> a_{2k},
\end{array}
\label{co1}
\ee
\be
f_{2k+1}(r)=
\Bigg\{
\begin{array}{c}
N_e^{-1}[N_e(r)+2k],\; r\leq a_{N-2k} \\
N_e^{-1}[2N-N_e(r)-2k],\; r> a_{N-2k},
\end{array}
\label{co2}
\ee
where $a_i=N_e^{-1}(i)$ and the integer $k$ runs for odd $N$
from $1$ to $(N-1)/2$, and for even $N$ from $1$ to
$(N-2)/2$.
For even $N$, we need an additional function
\be
f_N(r)=N_e^{-1}[N-N_e(r)].
\label{co3}
\ee
Equations~(\ref{co1})-(\ref{co3}) determine the distances $f_i(r)$ from the center of the remaining $N-1$ electrons when, say, electron 1 is at a distance $r$ from the center.
The relative angles between the electrons, leading to the vectorial co-motion functions $\vf_i(r)$, are determined by numerical minimization of Eq.~(\ref{eq_VeeSCE}) for each $r$. An example of such calculation for a 2D electron droplet of a uniform density with $N=4$ electrons (see Sec.~\ref{sec_2D} for more details) is shown in Fig.~\ref{fig_positionsD2N4}: As the position of the first electron changes on the $x$-axis from $x=0$ to $x=a_1=N_e^{-1}(1)$, the position of the second electron changes along the boldface curve from $r=a_2$ to $r=a_1$, the one of the third electron  from $r=a_2$ to $r=a_3$ and the one of the fourth electron from $r=a_4$ to $r=a_3$. The co-motion functions satisfy group properties such that the resulting electron-electron repulsion energy is invariant under exchange of two or more electrons, ensuring the indistinguishability of particles.\cite{SeiGorSav-PRA-07,GorSei-PCCP-10}

\section{Uniform electron droplets}

\subsection{Three dimensions}

We consider a homogeneous $N$-electron droplet with 
a constant density $n$ and a radius $R$. The density
can then be expressed simply as
\be
n(r)=
\Bigg\{
\begin{array}{c}
\frac{3N}{4\pi R^3},\; r\leq R \\
0,\; r>R.
\end{array}
\label{density}
\ee
In a physical sense, this density and the corresponding
wavefunction must be considered as limit cases (see
the end of this section). The expected number of 
electrons between $0$ and $r$ is now
\be
N_e(r) = N\left(\frac{r}{R}\right)^3\theta(R-r)+N\theta(r-R),
\ee
where $\theta$ is the Heaviside step function. 
The Hartree energy becomes now
\be\label{hartree}
U[n] = \frac{1}{2}\int d\vr' \,\int d\vr \,\frac{n(\vr)n(\vr')}{|\vr-\vr'|} = \frac{3N^2}{5R},
\ee
and the LDA exchange energy [Eq.~(\ref{LDA})] is
\be
E_x^{\rm LDA} = -\frac{3^{5/3}N^{4/3}}{2^{8/3}\pi^{2/3}R}.
\ee
For $N=2$, we can readily test the accuracy of the LDA with respect
to the exact exchange energy, $E_x^{\rm exact}(N=2) = -U(N=2)/2$.
We find $E_x^{\rm LDA}/E_x^{\rm exact}\approx 0.9621$. This ratio is
expected to approach unity as $N\rightarrow \infty$.

The co-motion functions [Eqs.~(\ref{co1})-(\ref{co3})] become
\be
f_{2k}(r)= \left|\frac{2k}{N}R^3-r^3\right|^{1/3},
\label{co3d1}
\ee
\begin{multline}
f_{2k+1}(r)= \\
\Bigg\{
\begin{array}{c}
\left(\frac{2k}{N}R^3+r^3\right)^{1/3},\; r\leq R(1-2k/N)^{1/3} \\
\left[\left(2-\frac{2k}{N}\right)R^3-r^3\right]^{1/3},\; r> R(1-2k/N)^{1/3},
\end{array}
\label{co3d2}
\end{multline}
and for even $N$ we have to add the last function $f_N(r)=|R^3-r^3|^{1/3}$.
These co-motion functions keep the electrons in different spherical 
shells (each one containing, in the quantum mechanical problem, on average one electron -- see the example of Fig.~\ref{fig_positionsD2N4}), while keeping the first derivative of the external potential
continuous.~\cite{SeiGorSav-PRA-07} The expectation value of the e-e interaction
in the strong-interaction limit can now be calculated from
\begin{multline}
\big<{\hat V}^\infty_{ee}(R)\big> = \\
4\pi\int_0^{a_1} dr\,r^2 n(r) V_{ee}(r,f_2(r),\ldots f_N(r),\Omega(r);R) \nonumber= \\
\frac{3N}{R^3}\int_0^{RN^{-1/3}} dr\,r^2 V_{ee}(r,f_2(r),\ldots f_N(r),\Omega(r);R),
\end{multline}
where we have used the fact that integrating between $0$ and $R$ is equivalent\cite{SeiGorSav-PRA-07}
to integrating $N$ times between $0$ and $a_1$, where $a_1=N_e^{-1}(1)=R N^{-1/3}$. The function $\Omega(r)$ denotes all the relative angles between
the electrons as a function of $r$ and is calculated numerically.\cite{SeiGorSav-PRA-07}
Changing variables $x=r/R$ leads to
\begin{multline}
\big<{\hat V}^\infty_{ee}(R)\big> = \\
3N\int_0^{N^{-1/3}} dx\,x^2 V_{ee}(x,f_2(x),\ldots f_N(x),\Omega(x);R).
\end{multline}
Upon coordinate scaling it is easy to see that 
\begin{multline}
V_{ee}(x,f_2(x),\ldots f_N(x),\Omega(x);R)= \\
\frac{1}{R} V_{ee}(x,f_2(x),\ldots f_N(x),\Omega(x);R=1).
\end{multline}
Finally, we can write Eq.~(\ref{eq_Lambda}) as a {\em function} of $N$,
\be
\Lambda_3(N) = \frac{\big<{\hat V}^\infty_{ee}(R=1)\big> - 3N^2/5}{-3^{5/3}\pi^{-2/3}(N/4)^{4/3}},
\ee
where, as said, $\big<{\hat V}_{ee}(R=1)\big>$ is calculated numerically.
Special cases are $N=1$ and $N=2$ yielding analytic expressions. For a single
electron
$\big<{\hat V}_{ee}\big>$ is trivially zero and we find
$\Lambda_3(N=1)=4(2\pi/3)^{2/3}/5\approx 1.310$. 
For $N=2$ we get
$\big<{\hat V}_{ee}(R=1)\big>=3\left(8-2^{1/3}\,\Gamma(1/6)\Gamma(4/3)/\sqrt{\pi}\right)/20$
leading to $\Lambda_3(N=2)\approx 1.498$. Both values are lower than those given for $\tilde{\lambda}_3(N=1)$ and $\tilde{\lambda}_3(N=2)$
in Ref.~\onlinecite{OdaCapTri-JCTC-09}.

\begin{figure}
\includegraphics[width=0.9\columnwidth]{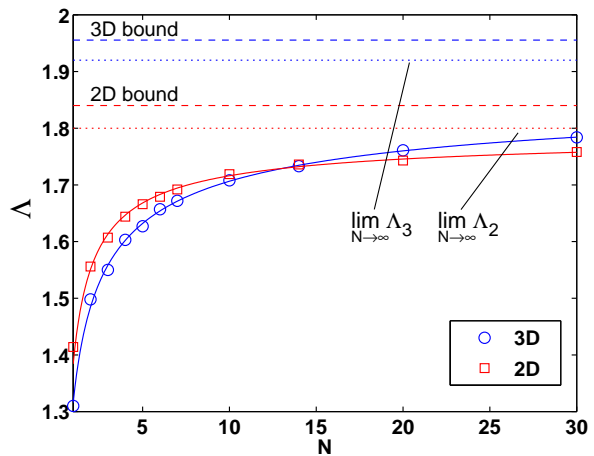}
\caption{(Color online) Values obtained for $\Lambda_3$ (circles) 
and $\Lambda_2$ (squares) as a function of the electron number $N$.
The dotted lines show the estimated limit values for uniform 
electron droplets when $N$ goes to infinity. The dashed lines
show the bounds ${\bar \lambda}_3$ and ${\bar \lambda}_2$ for 
three- and two-dimensional systems, respectively,
corresponding to the low-density limit of the homogeneous
electron gas.~\cite{RasPitCapPro-PRL-09}}
\label{3d2d}
\end{figure}

Our numerical results for larger $N$ are summarized in Table~\ref{table}.
\begin{table}
  \caption{\label{table} 
Calculated values for $\Lambda_3$ and $\Lambda_2$ 
as a function of the number of electrons $N$ in uniform
strictly correlated electron droplets in three and two dimensions, 
respectively.
}
  \begin{tabular*}{0.5\columnwidth}{@{\extracolsep{\fill}} c | c c }
  \hline
  \hline
  $N$    & $\Lambda_3$ & $\Lambda_2$ \\
 \hline
1   &  1.310   & 1.414 \\
2   &  1.498   & 1.556   \\
3   &  1.550   & 1.607   \\
4   &  1.603   & 1.644   \\
5   &  1.627   & 1.666   \\
6   &  1.657   & 1.679   \\
7   &  1.672   & 1.692   \\
10   & 1.708   & 1.719   \\
14   & 1.733   & 1.736   \\
20   & 1.761   & 1.743  \\
30   & 1.784   & 1.758\\
  \hline
  \hline
  \end{tabular*}
\end{table}
The results are also plotted (as circles) in Fig.~\ref{3d2d}.
The curve intersecting the tabulated values has been obtained 
by numerical fitting of $\langle{\hat V}^\infty_{ee}(R=1)\rangle$ using a liquid-drop model expansion, which leads for $\Lambda_3(N)$ to the formula
\be
\Lambda^{\rm fit}_3(N) = -\frac{4}{3}\left(\frac{2\pi}{3}\right)^{2/3}\left(a_1+a_2\,N^{-1/3}+a_3\,N^{-2/3}\right),
\ee
where $a_1=-0.879717$, $a_2=0.153634$, and $a_3=0.123195$. 
When $N$ goes to infinity, the fit yields a value 
$\Lambda_3(N\rightarrow\infty)\approx 1.92$ plotted as a dotted line 
in Fig.~\ref{3d2d}. This value is lower than the 
bound proposed in Ref.~\onlinecite{RasPitCapPro-PRL-09} for
3D systems corresponding to ${\bar \lambda}_3=1.9555$ (dashed line).
However, the difference is rather small ($\sim 2\%$) and it is actually 
within the error associated to the fitting procedure. Our values for 
$\Lambda_3(N)$ as well as our fitting curve are always below the model 
for $\tilde{\lambda}_3(N)$ proposed by Odashima {\it et al.}\cite{OdaCapTri-JCTC-09}

It should be noted that, {\em per se}, the density
given in Eq.~(\ref{density}) corresponds to a non-differentiable
wavefunction due to the sharp edge at $r=R$. Therefore,
it is important to examine whether the results above are valid
when considering the density as a limit case of a physical 
density. A simple choice would be a density profile of the
form of a Fermi function, i.e.,
\be
{\tilde n}(r)=\frac{{\rm const}}{e^{\alpha(r-R)}+1},
\ee
where the numerator is the normalization constant
and the value for $\alpha$ determines the sharpness 
of the edge. The limit $\alpha\rightarrow\infty$ corresponds
to the density of the form of Eq.~(\ref{density}). 
\begin{figure}
\includegraphics[width=0.8\columnwidth]{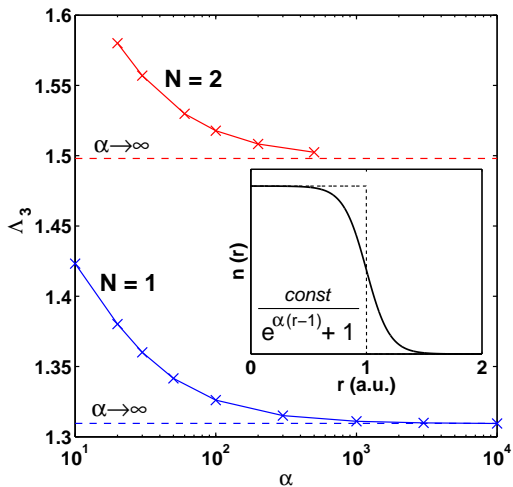}
\caption{(Color online) 
Values obtained for $\Lambda_3$ with $N=1$ (bottom) and $N=2$ (up)
when a soft tail of the density (solid line in the inset) is
applied. In the limit of a sharp edge (dashed line in
the inset), corresponding to $\alpha\rightarrow\infty$,
the results for the proposed homogeneous density droplet in
Eq.~(\ref{density}) are reproduced (dashed lines).
}
\label{fermi}
\end{figure}
Figure~(\ref{fermi}) shows the values obtained numerically 
for $\Lambda_3$ with $N=1$ (bottom) and $N=2$ (up)
as a function of $\alpha$. From the figure it is clear
that the (analytic) values corresponding to the original
(sharp) density are approached as $\alpha\rightarrow\infty$;
this is particularly convincing with $N=1$ where we can numerically
study very large values for $\alpha$ (even several orders of 
magnitude larger than those shown in the figure). Thus,
applying densities of the form of Eq.~(\ref{density}), can be seen as a limiting
case for a quantum system.

\subsection{Two dimensions}
\label{sec_2D}
In 2D a homogeneous $N$-electron droplet with a radius $R$
is defined by the density
\be
n(r)=
\Bigg\{
\begin{array}{c}
\frac{N}{\pi R^2},\; r\leq R \\
0,\; r>R,
\end{array}
\label{density2}
\ee
and the expected number of electrons between $0$ and $r$ is
\be
N_e(r) = N\left(\frac{r}{R}\right)^2\theta(R-r)+N\theta(r-R).
\ee
The Hartree energy is more tricky to calculate than in 3D, since using
Eq.~(\ref{hartree}) directly leads to an elliptic integral.
However, we can use the fact that for a 2D disk with a radius $r$ and
a constant density $n$, the Hartree potential {\em at the rim} of the
disk is $V_H=4\,n\, r.$\footnote{See, e.g., pp. 67 in M. Seidl, 
Habilitation thesis (University of Regensburg);
Eq. (2.26) in {\it Berkeley Physics Course}, (McGraw-Hill, New York, 1965);
pp. 41 in Ref.~\onlinecite{GiuVig-BOOK-05}; Eq. (13) in 
M. Seidl, J. P. Perdew, and M. Levy, Phys. Rev. A {\bf 59}, 51 (1999).} 
The Hartree energy is now
equal to the work required to charge the disk from the center ($r=0$)
to the rim ($r=R$),
\be
U[n] = \frac{N}{\pi R^2}2\pi\int_0^R dr\,r V_H(r) = \frac{8N^2}{3\pi R}.
\ee
The LDA exchange energy is
\be
E_x^{\rm LDA} = -\frac{2^{5/2}N^{3/2}}{3\pi R}.
\ee
Again, for the special case of $N=2$ we can compare $E_x^{\rm LDA}$
with the exact exchange energy, $E_x^{\rm exact}=-U/2$.
We find $E_x^{\rm LDA}/E_x^{\rm exact} = 1$, i.e.,
the LDA exchange energy is exact for $N=2$. 


In 2D the co-motion functions have the same form as in 3D 
[Eqs.~(\ref{co3d1}) and (\ref{co3d2})] apart from the
change in exponents as $3\rightarrow 2$ and $1/3\rightarrow 1/2$.
The expectation value of the e-e interaction operator 
in the strong-interaction limit becomes
\begin{multline}
\big<{\hat V}^\infty_{ee}(R)\big> = \\
2N\int_0^{N^{-1/2}} dx\,x\, V_{ee}(x,f_2(x),\ldots f_N(x),\Omega(x);R),
\end{multline}
and after scaling of the distances we find
\be
\Lambda_2(N) = \frac{\big<{\hat V}^\infty_{ee}(R=1)\big> - 8N^2/(3\pi)}{-2^{5/2}N^{3/2}/(3\pi)}.
\ee
For $N=1$ we get now simply $\Lambda_2(N=1)=\sqrt{2}$,
and the two-electron case yields $\Lambda_2(N=2)=2-3\pi\left\{8+\sqrt{2}\left[\ln 2+\ln(2-\sqrt{2})-3\ln(2+\sqrt{2})\right]\right\}\approx 1.556$.
Results for larger $N$ are given in Table~\ref{table}
and Fig.~\ref{3d2d}. Interestingly, the 2D values are higher than the 3D ones at small $N$, but at 
$N\sim 15$ they go below the 3D curve. Again, we use a liquid-drop-model expansion to fit our data, leading to
\be
\Lambda^{\rm fit}_2(N) = -\frac{3\pi}{4\sqrt{2}}\left(b_1+b_2\,N^{-1/2}+b_3\,N^{-1}\right),
\ee
with $b_1=-1.0814$, $b_2=0.121609$, and $b_3=0.129014$. 
The large-$N$ limit yields $\Lambda_2(N\rightarrow\infty)=1.80$, which
is, also in this case, $\sim 2\%$ lower than the 2D 
bound ${\bar \lambda}_2=1.84$ in Ref.~\onlinecite{RasPitCapPro-PRL-09}.

\subsection{One dimension}\label{1D}

\begin{figure}
\includegraphics[width=0.75\columnwidth]{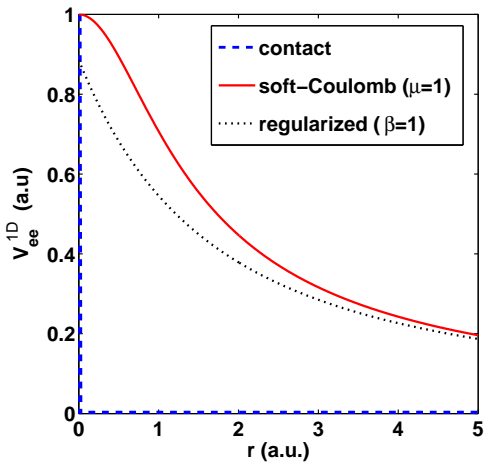}
\caption{(Color online) Different e-e interaction potentials
considered for one-dimensional systems.}
\label{1D_interaction}
\end{figure}

As mentioned in Sec.~\ref{lower}, the Coulomb interaction is
ill-defined in 1D. Various forms for physically reasonable
1D e-e interaction operators have been suggested, and below we  
focus on three of them: the contact, soft-Coulomb, and
regularized interaction.
The shapes of the interaction potentials are visualized 
in Fig.~\ref{1D_interaction}.

Regardless of the type of e-e interaction, the 1D homogeneous
electron droplet has a density
\be
n(x)=
\Bigg\{
\begin{array}{c}
\frac{N}{2 R},\; |x|\leq R \\
0,\; |x|>R,
\end{array}
\label{density1}
\ee
and the expected number of electrons between $-\infty$ and $x$
(here corresponding to the cumulative distribution function) is
\be
N_e(x)=\int_{-\infty}^x\,dx'\,n(x')=
\Bigg\{
\begin{array}{c}
0,\; x<-R \\
\frac{N}{2 R}x+\frac{N}{2},\; |x|\leq R \\
N,\; x>R,
\end{array}
\ee
The co-motion functions $f_i(x)$ can be found in a straightforward fashion.
When the first electron ($i=1$) is set at $-R$, the electron
$i$ is located at $a_{i-1}=2(i-1)R/N - R$. This leads to
\be
f_i(x)=
\Bigg\{
\begin{array}{c}
N_e^{-1}[N_e(x)+i-1],\; x\leq N_e^{-1}(N+1-i) \\
N_e^{-1}[N_e(x)-(N+1-i)],\; x> N_e^{-1}(N+1-i),
\end{array} 
\ee
and after substituting $N_e^{-1}$ we get
\be
f_i(x)=
\Bigg\{
\begin{array}{c}
x+2\frac{R}{N}(i-1),   \; x\leq 2\frac{R}{N}(1-i)+R \\
x+2\frac{R}{N}(i-1)-2R,\; x\leq 2\frac{R}{N}(1-i)+R.
\end{array}
\ee

The expectation value of the e-e interaction in
the strong-interaction limit can be written as
\bea
\big<{\hat V}^\infty_{ee}(R)\big> & = & \int_{-\infty}^\infty dx\,
\frac{n(x)}{N}\sum_{i>j}V_{ee}\left(|f_i(x)-f_j(x)|\right) \nonumber \\
& = & \sum_{i>j}V_{ee}\left(\left|2\frac{R}{N}(i-j)\right|\right),
\label{vee_1d}
\eea
where we have replaced the full integral by $N$ integrals between
$-R$ and $a_1=-R+2R/N$, so that the difference between two co-motion
functions is always $f_i(x)-f_j(x)=2(i-j)R/N$. Thus, we notice that
the strong-interaction limit in 1D, in the case of a {\em uniform} 
density, corresponds to the Wigner-crystal solution due to the 
translational invariance. In 3D and 2D instead, the distances 
between $f_i(r)$ always depend on $r$.

Next, let us write the Hartree energy and the LDA exchange energy in a general form,
\begin{eqnarray}\label{hartree_1d}
U[n] & = & \frac{N^2}{8R^2}\int_{-R}^R\int_{-R}^R dx\,dx'\,V_{ee}(|x-x'|), \\
\label{lda_1d}
E_x^{\rm LDA}  &  =  & \int_{-\infty}^\infty dx\,n(x)\epsilon_x(n),
\end{eqnarray}
where the exchange energy per electron in a 1D HEG  
is~\cite{GiuVig-BOOK-05}
\be
\epsilon_x(n)=-\frac{1}{2\pi}\int_0^{2k_F} dq\,{\tilde V}_{ee}(q)\left(1-\frac{q}{2k_F}\right),
\ee
${\tilde V}_{ee}$ is the Fourier transform of the e-e interaction and 
$k_F=\pi n/2$ is the Fermi vector in 1D.
In the following, we will use the above general expressions
to compute $\Lambda_1$ in Eq.~(\ref{eq_Lambda}) as a function of $N$.

\subsubsection{Contact interaction}

The contact (or delta) interaction is defined as
\be
V_{ee}(x) = \eta\,\delta(x)
\ee
(see the dashed line in Fig.~\ref{1D_interaction})
and its Fourier transform is simply ${\tilde V}_{ee}(q)=\eta$,
where $\eta$ is a dimensionless constant. The Hartree
energy [Eq.~(\ref{hartree_1d})] becomes
\be
U[n]=\frac{\eta N^2}{4R},
\ee
and the LDA exchange energy [Eq.~(\ref{lda_1d})] is
\be
E_x^{\rm LDA}=-\eta\frac{N^2}{8R}.
\ee

We see that, in this case, the calculation of the constrained minimization of Eq.~(\ref{eq_theSCEfunctional}) does not have a unique minimizing wavefunction. Indeed, with the contact interaction the minimum value $\langle\hat{V}_{ee}^\infty\rangle=0$ can be produced with any wavefunction that prevents the electrons to be at the same position while yielding the assigned density. The strictly-correlated wavefunction is just one of those. 
Thus, we trivially obtain
$\Lambda_1=2$, which is {\em independent of N}, and coincides with the lower bound ${\bar \lambda}_1$
of Ref.~\onlinecite{RasPitCapPro-PRL-09}.
 
\subsubsection{Soft-Coulomb interaction}

The soft-Coulomb interaction is defined as
\be
V_{ee}(x) = \frac{1}{\sqrt{x^2+\mu^2}},
\ee
where $\mu$ is the softening (or cutoff) parameter.
The potential is visualized as a solid line in Fig.~\ref{1D_interaction}.
Its Fourier transform is ${\tilde V}_{ee}(q)=2\,K_0(\mu q)$,
where $K_0$ is the modified Bessel function of the second kind.\cite{AbrSte-BOOK-65}
This expression leads to the LDA exchange energy of the form
\bea \label{log}
E_x^{\rm LDA} & = & -\frac{N^2}{4R}\int^1_0 dx\,2(1-x)\,K_0\left(\frac{\pi\mu N}{2R}x\right) \\
& = & -\frac{N^2}{4R}\left[\ln\left(\frac{4R}{\pi\mu N}\right)+\frac{3}{2}-\gamma\right]+\mathcal{O}\left(\frac{\pi\mu N}{2R}\right)^2, \nonumber
\eea
where $\gamma\approx 0.577$ is Euler's constant. The leading term at small $\pi\mu N/(2R)$ 
agrees with the result of Fogler.~\cite{Fog-PRL-05}

The calculation of the Hartree energy leads to a tedious integral
but finally yields an analytic expression,
\bea
U[n] & = & \frac{N^2}{4R}\Bigg\{\mu/R-\sqrt{4+(\mu/R)^2}-\ln(\mu/R) \nonumber \\
& + & \frac{1}{2}\ln\left[ \frac{\left(\sqrt{4+(\mu/R)^2}+2\right)^3} {\sqrt{4+(\mu/R)^2}-2}\right]\Bigg\} \nonumber \\
& = & \frac{N^2}{4R} \Bigg\{ 4\ln(2) -2-2\ln\left(\frac{\mu}{R}\right)+\frac{\mu}{R}-\frac{1}{8}\left(\frac{\mu}{R}\right)^2 \nonumber \\
& + & \frac{1}{256}\left(\frac{\mu}{R}\right)^4 - \frac{1}{3072}\left(\frac{\mu}{R}\right)^6 + \mathcal{O}\left(\frac{\mu}{R}\right)^8 \Bigg\},
\eea
where we also give the series expansion for small values of $(\mu/R)$, which is the regime of 
our primary interest (see below).
Finally, the strong-interaction limit for the e-e interaction in Eq.~(\ref{vee_1d}) leads to
\be
\big<{\hat V}^\infty_{ee}(R)\big> = \frac{N}{R}\sum_{i>j}^N\frac{1}{\sqrt{4(i-j)^2+\left(\frac{\mu N}{R}\right)^2}}.
\ee

\begin{figure}
\includegraphics[width=0.7\columnwidth]{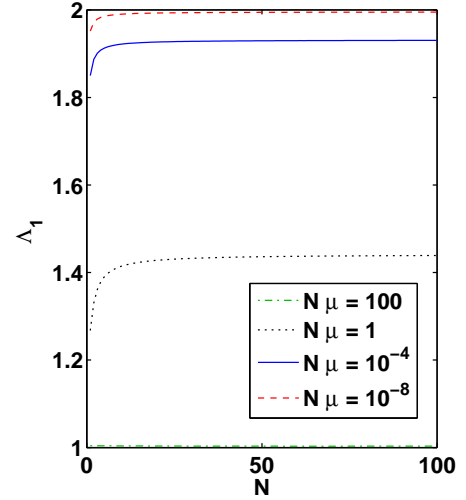}
\caption{(Color online) Values obtained for $\Lambda_1$ as a function 
of the electron number $N$
with different values for $N\mu$, where $\mu$ is the softening parameter
in the soft-Coulomb interaction.}
\label{soft}
\end{figure}

Similarly to the 3D and 2D case, we may set $R=1$. Thus, values for $\Lambda_1$ 
essentially depend on $N$ and $\mu$, and in $E_x^{\rm LDA}$ and $\big<{\hat V}_{ee}^\infty(R)\big>$ also
through their product $N\mu$. Hence, in the following we fix $N\mu$ and
examine numerically the behavior of $\Lambda_1$ as a function of $N$.
As visualized in Fig.~\ref{soft},
we find that increasing values for $N\mu$ lead to a decrease in $\Lambda_1$,
whereas decreasing $N\mu$ leads to an asymptotic approach of $\Lambda_1$ toward two.
This tendency is in agreement with the bound ${\bar \lambda}_1=2$
in Ref.~\onlinecite{RasPitCapPro-PRL-09}, 
where it was assumed that the Lieb-Oxford-like bound 
in a {\em soft-Coulombic} 1D system
has the same general form as Eq.~(\ref{bound}) upon the multiplication of the 
logarithmic factor in Eq.~(\ref{log}).

\subsubsection{Regularized interaction}

As the third alternative for the e-e interaction in 1D we consider
the regularized form of the Coulomb interaction in 1D. In particular,
we take the representation of the Yukawa interaction in an 
infinite cylindrical wire of 
radius $\beta$ (Ref.~\onlinecite{GiuVig-BOOK-05}). The system is then
studied in the limit where the (finite) range of this interaction
is larger than any other length scale except the length of the wire.
The resulting interaction potential in
the momentum space is\cite{GiuVig-BOOK-05}
\be
{\tilde V}_{ee}(q)=e^{\beta^2 q^2}\,{\rm Ei}\,(\beta^2 q^2),
\ee
where ${\rm Ei}(z)$ is the exponential-integral function.\cite{AbrSte-BOOK-65} In real space
the interaction can be written as
\be
V_{ee}(x)=\frac{\sqrt{\pi}}{2 \beta}\,e^{x^2/(4\beta^2)}\,{\rm erfc}\left(\frac{|x|}{2\beta}\right),
\ee
where ${\rm erfc}(z)=1-{\rm erf}(z)$ is the complementary error function. The
potential is visualized as a dotted line in Fig.~\ref{1D_interaction}.

As can be expected, all terms required to calculate $\Lambda_1$ become
now rather cumbersome. The LDA exchange energy is
\be\label{exreg}
E_x^{\rm LDA}=-\frac{N^2}{4R}\,\int_0^1 dq\,e^{q^2 b^2}\,{\rm Ei}(-q^2 b^2)(1-q),
\ee
where $b=\pi\beta N/(2R)$. 
The Hartree integral is more straightforward to calculate 
in Fourier space. This leads to
\bea \label{ureg}
U[n] & = & \frac{1}{2\pi}\int_0^\infty dq\,{\tilde n}^2(q)\,{\tilde V}_{ee}(q) \nonumber \\
     & = & \frac{N}{2\pi}\int_0^\infty dq\,\frac{\sin^2(q R)}{(q R)^2}\,{\tilde V}_{ee}(q) \nonumber \\
     & = & \frac{\pi^{3/2}}{2}\Bigg[\frac{\beta}{R}-\frac{\beta}{R}e^{R^2/\beta^2}+\sqrt{\pi}\,{\rm erfi}\left(\frac{R}{\beta}\right)\Bigg]  \nonumber \\
     & - & \frac{1}{3}\frac{R^2}{\beta^2}\, {}_p F_q\left(1,1;2,\frac{5}{2};\frac{R^2}{\beta^2}\right),
\eea
where erfi is the imaginary error function and ${}_p F_q$ 
is the generalized hypergeometric function.\cite{AbrSte-BOOK-65}
The strong-interaction limit of the e-e interaction becomes
\be
\big<{\hat V}^\infty_{ee}(R)\big> = \frac{\sqrt{\pi}}{2\beta}\sum_{i>j}^N{\rm Exp}\left[\frac{R\,(i-j)}{\beta N}\right]^2
{\rm erfc}\,\left(\frac{R|i-j|}{\beta N}\right).
\ee

Now, to calculate $\Lambda_1$ for the regularized interaction using the 
quantities above, we have to restrict the parameter range. First,
$E_x^{\rm LDA}$ in Eq.~(\ref{exreg}) is unstable for small $b$, and
$U[n]$ in Eq.~(\ref{ureg}) is unstable for small $\beta/R$. Therefore,
in both cases we numerically compute the series expansions up to the
second order of these quantities. This corresponds to the physically
justified small-$\beta$ limit of the infinite cylinder. 
Second, $\big<{\hat V}^\infty_{ee}(R)\big>$ is unstable for large $R|i-j|/(\beta N)$,
so we again use the series expansion. 

\begin{figure}
\includegraphics[width=0.7\columnwidth]{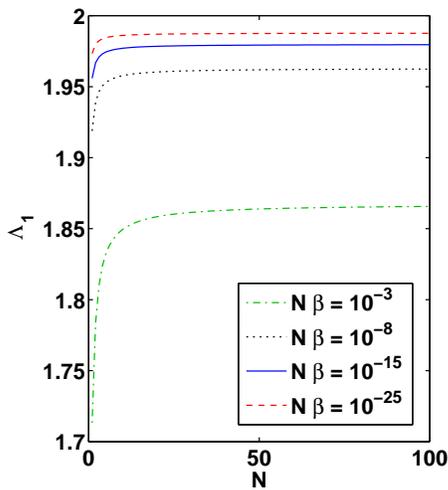}
\caption{(Color online) Values obtained for $\Lambda_1$ as a function 
of the electron number $N$
with different values for $N\beta$, where $\beta$ is the cutoff 
parameter in the regularized electron-electron interaction.}
\label{reg}
\end{figure}

In Fig.~\ref{reg}
we show the behavior of $\Lambda_1$ as a function of $N$
with different (small) values of $N\beta$. As in the soft-Coulombic
case, decrease in the ``cutoff'' parameter in the e-e interaction
leads to a tendency toward $\Lambda_1=2$, although
in this case the approach is very slow as a function of $N\beta$.
Nevertheless, the results are in line with other types of
e-e interaction. It should be noted that for this regularized
interaction a Lieb-Oxford-like bound has {\em not} been constructed
or even proposed before. Therefore, on the basis of our results
above we may suggest $\bar{\lambda}_1=2$ for the bound, thus agreeing
with 1D systems interacting through contact or soft-Coulomb
interaction. 

\section{Conclusions and perspectives}

To summarize, we have examined the properties of
strictly correlated electron droplets having a 
locally uniform density.
In particular, we have used the theory of strictly
correlated electrons to test the validity of
the lower bounds proposed in Ref.~\onlinecite{RasPitCapPro-PRL-09} on the indirect 
electron-electron interaction in $D$-dimensional
quantum systems. We have found that the bound
is satisfied in all dimensions, although  it
is approached as a function of the electron number.
In one-dimensional droplets the bound is
reached asymptotically regardless of the type
of electron-electron interaction considered in this
work. In two and three dimensions, 
we obtain values being a few percent above the
lower bounds. 

Our results 
can be taken as useful reference data for
future investigations of strongly correlated
systems in general, as well as for the development
and testing of exchange-correlation functionals
in the framework of density-functional theory. 
In future work we plan to investigate  different densities, trying to 
challenge the bound maximally, as well as trying to provide 
reference data to construct $N$-dependent bounds.~\cite{OdaCapTri-JCTC-09}

\begin{acknowledgments}
This work was supported by the Academy of
Finland and by the Netherlands Organization for Scientific Research (NWO) through a Vidi grant.
\end{acknowledgments}


\end{document}